\documentclass[prodmode,acmcsur]{acmsmall} 

\usepackage{booktabs} 
\usepackage[ruled]{algorithm2e}

\SetAlFnt{\small}
\SetAlCapFnt{\small}
\SetAlCapNameFnt{\small}
\SetAlCapHSkip{0pt}
\IncMargin{-\parindent}

\acmVolume{0}
\acmNumber{0}
\acmArticle{1}
\acmYear{2014}
\acmMonth{0}

\usepackage{listings}
\usepackage{rotating}

\begin{document}
%
\title{A Pattern-based Survey and Categorization of Network Covert Channel Techniques}



%
\markboth{S. Wendzel, S. Zander et al.}{A Pattern-based Survey and Categorization of Network Covert Channel Techniques}

\title{A Pattern-based Survey and Categorization of Network Covert Channel Techniques}
\author{STEFFEN WENDZEL
\affil{Fraunhofer Inst. for Communication, Inf. Processing and Ergonomics FKIE}
SEBASTIAN ZANDER
\affil{Centre for Advanced Internet Architectures, Swinburne Univ. of Technology}
BERNHARD FECHNER
\affil{Department of Systems and Networking, Univ. of Augsburg}
CHRISTIAN HERDIN
\affil{Department of Computer Science, Univ. of Rostock}
}


\newcommand{\NumOfAnalyzedTechniques}{109}
\newcommand{\NumOfPatterns}{11}  
\newcommand{\ProzentOfPatternsInTopCategories}{69.7}

\renewcommand{\labelenumi}{\arabic{enumi}. }

\begin{abstract}

Network covert channels are used to hide communication inside network protocols. Within the last decades, various techniques for covert channels arose. 
We surveyed and analyzed \NumOfAnalyzedTechniques{} techniques developed between 1987 and 2013 and show that these techniques can be reduced to only \NumOfPatterns{} different patterns. Moreover, the majority (\ProzentOfPatternsInTopCategories\%) of techniques can be categorized in only four different patterns, i.e. most of the techniques we surveyed are very similar.
We represent the patterns in a hierarchical catalog using a pattern language.
Our pattern catalog will serve as a base for future covert channel novelty evaluation. 
Furthermore, we apply the concept of pattern variations to network covert channels. With pattern variations, the context of a pattern can change. For example, a channel developed for IPv4 can automatically be adapted to other network protocols.
We also propose the pattern-based covert channel optimizations pattern hopping and pattern combination. 
Finally, we lay the foundation for pattern-based countermeasures: While many current countermeasures were developed for specific
channels, a pattern-oriented approach allows to apply one countermeasure to multiple channels. Hence, future countermeasure
development can focus on patterns, and the development of real-world protection against covert channels is greatly simplified.
\end{abstract}





\acmformat{Steffen Wendzel, Sebastian Zander, Bernhard Fechner and Christian Herdin. Pattern-based Categorization of Network Covert Channel Techniques}

\begin{bottomstuff}
Author's addresses: S. Wendzel, Fraunhofer Institute for Communication, Information Processing and Ergonomics FKIE, Bonn, Germany; S. Zander, Centre for Advanced Internet Architectures, Swinburne University of Technology, Melbourne, Australia; B. Fechner, University of Augsburg, Department of Systems and Networking, Augsburg, Germany; C. Herdin, University of Rostock, Department of Computer Science, Rostock, Germany
\end{bottomstuff}

\maketitle

\section{Introduction}

Covert channels represent unforeseen communication methods that break security policies \cite{DBLP:journals/cacm/Lampson73}. \emph{Network} covert channels transfer information through networks in ways that hide the fact that communication takes place (hidden information transfer). They are used in scenarios where normal communication is too revealing and just using encryption is not sufficient.

Network covert channels are considered a \emph{serious threat to Internet users} \cite{gianvecchio07}, because they can be used to hide command and control traffic of botnets, to coordinate DDoS attacks, to hide military and secret service communications, and to secretly leak sensitive data  \cite{springerlink:10.1007/11767831_10,journals/comsur/ZanderAB07}. On the other hand, network covert channels are a dual-use good and can prevent illicit information transferred by journalists or whistle-blowers from being detected, and thus support the freedom of speech in networks with censorship \cite{journals/comsur/ZanderAB07}.

Today's hiding techniques embed hidden information either
\begin{enumerate}
\item in \emph{protocol data units} (PDUs), e.g. in unused or reserved header elements (sometimes also called \emph{header fields}),
\item or through the timing of PDUs or protocol commands, e.g. by encoding a hidden message as a sequence of inter-arrival times or as a manipulated packet order.
\end{enumerate}

A large amount of research was accomplished within the last decades to evaluate attributes of network protocols (e.g. IPv4, IPv6, TCP, and HTTP) regarding their potential to hide information. On the other hand, only little work exists on providing a general, protocol-independent approach. Although coarse categorizations of network covert channel techniques exist \cite{journals/comsur/ZanderAB07,CC:ContextBasedView,OptimizationofCCIdentification,Llamas:Survey,EntropyTaxonomyNCC}, no comprehensive and current catalog of the existing techniques is available.

Moreover, current techniques to counter network covert channels focus on single covert channels instead of common characteristics of multiple channels. The combination of dozens of countermeasures is required to achieve an acceptable protection, which is problematic in practice.

We consider a taxonomy for covert channel techniques very important in order to provide a framework to classify current and future research in the field, to determine similarities between techniques and to streamline the identification of novel countermeasures.

\emph{Patterns} are a universal technique, which can be used to create taxonomies in a generic manner \cite{CHI2003}. In particular, the \emph{Pattern Language Markup Language} (PLML) provides a consistent formalization of pattern descriptions and is the standard pattern language in the human computer interaction field \cite{CHI2003}.


We apply the approach of pattern languages to network covert channels, extract common patterns for hiding techniques and combine them in a novel hierarchy. In comparison to existing taxonomy approaches, we also cover recent covert channels from 2009 to 2013. The focus of our pattern catalog is less on technical aspects but on the common abstract behavior of covert channel techniques, which is also a difference to existing categorizations.

We describe the identified covert channel patterns using an extensible PLML-based pattern catalog. Our catalog simplifies the future classification and novelty evaluation of upcoming covert channels. Only hiding techniques which require the integration of a new pattern into the catalog are very novel, others are simply variations of existing patterns. We show that the surveyed techniques can be reduced to only \NumOfPatterns{} different patterns. Moreover, the majority (\ProzentOfPatternsInTopCategories\%) of techniques can be categorized in only four different patterns, i.e. most of the covert channel techniques we surveyed are very similar. Furthermore, our pattern catalog represents a systematic approach for identifying network covert channels in protocols in order to overcome the problem of requiring an \emph{exhaustive search} \cite{SilenceInLANs}.

In addition, we present the idea of pattern \emph{variation}. Pattern variation is based on pattern \emph{transformation} \cite{PatternTransformation,Engel:PatternTransNEU}, which allows authors and developers to alter the existing \emph{context} of a pattern. For instance, a desktop browser interface pattern can be transformed to a user interface pattern for mobile devices and vice versa, i.e. the context changes from desktop to a mobile device \cite{PatternTransformation}.

Pattern variation is the first transformation-like approach for covert channels. We define the utilized network protocol as the pattern's context. Thus, a pattern's application can change from one network protocol to another -- without re-implementing the hiding technique itself.

We also explain the improvement of pattern-based covert channels by introducing the concepts of \emph{pattern combination} and \emph{pattern hopping}. Pattern combination allows to use multiple patterns at the same time (e.g. for a single network packet or frame) to increase throughput while pattern hopping randomizes the use of patterns over time to increase stealthiness.

Furthermore, we motivate the development of countermeasures for network covert channels based on patterns. With patterns, covert channel protection in practice will become more realistic as the number of required countermeasures can be reduced greatly by targeting hiding techniques represented through generic patterns instead of aiming at specific hiding techniques.

The remainder of this article is structured as follows. Section~\ref{Sect:FundamRelWorkCC} covers fundamentals of network covert channels and discusses previous taxonomies while Section~\ref{Sect:FundamRelWorkPatterns} explains the concept of patterns and their use in our taxonomy. We introduce the identified covert channel patterns and our hierarchical pattern catalog in Section~\ref{Sect:CCPatterns} and present our concept of pattern variation in Section~\ref{Sect:Transformation}. Section~\ref{Sect:PreventionDetection} motivates and discusses pattern-based countermeasures. A conclusion follows in Section~\ref{Sect:Concl}.

\section{Covert Channel Fundamentals and Related Work}\label{Sect:FundamRelWorkCC}

First, we provide background information on network covert channels and afterwards discuss previous work on the categorization of covert channels.

\subsection{Covert Channels}

This section provides some background on covert channels and introduces the terminology used in the rest of the article.

\subsubsection{History}
Covert channels were introduced by Lampson in 1973 and represent a technique for security policy-breaking communication that was not foreseen by a system designer \cite{DBLP:journals/cacm/Lampson73,MurdochPhDThesis}. Covert channels became an important  topic in the military context where a high-security process having sensitive information (the HIGH process) must be prevented from leaking information to a process with lower security (the LOW process) through covert channels \cite{CC_Elimination_Protos}. For instance, a LOW process classified as ``SECRET'' should not be capable to access ``TOP SECRET'' data from a HIGH process.

Initially, the research community focussed on \emph{local covert channels} -- covert channels that can be used to leak data from a HIGH process to a LOW process on the \emph{same} system. With the rise of computer networks starting from the early 1990's, the focus shifted towards \emph{network covert channels}. Network covert channels encode hidden data in network protocols (also referred to as  \emph{overt channels}). For network covert channels, the focus is not solely on  establishing a policy-breaking communication anymore. They are more broadly viewed as approaches to provide a \emph{hidden} communication channel \cite{Millen:20Years}.
Traditionally, network covert channels were classified as \emph{storage channels}, which encode hidden data in protocol fields, and \emph{timing channels}, which hide information by manipulating the timing of frames, packets, or messages.

Fisk \emph{et al.} distinguish between unstructured and structured carriers for hiding techniques \cite{FiskEtAl03}. Unstructurred carriers are human-interpretable and placed into the payload of network packets (e.g. audio or video streaming content). Our work concentrates on structured, i.e. machine-interpretable, carriers such as network protocol headers. 

\subsubsection{Adversary Scenario}

Simmons introduced the so-called \emph{Prisoner's Problem} in 1983 \cite{DBLP:conf/crypto/Simmons83}. Alice and Bob are kept in prison cells, separated from each other, and want to cooperate in order to escape from jail. The only way for Alice and Bob to exchange messages is to give the  messages to the warden Walter. Walter can read all messages exchanged between both prisoners. He decides whether he delivers or manipulates a message and can even forge messages. Alice and Bob therefore need to use a hidden communication which Walter is not aware of, i.e. a covert channel, to achieve their goal. For instance, the prisoners could alter the row pitch between written lines in letters they hand to the warden.

In the context of covert channels, an adversary is either active or passive.
A passive adversary, or \emph{passive warden}, tries to determine the presence of a network covert channel and to extract the embedded message or to prove the involvement of a party in the covert communication \cite{IH:Terminology}. An active adversary, or \emph{active warden}, can also try to modify the covert communication by removing or blocking elements within the data transfer or inserting its own (bogus) messages into the channel \cite{IH:Terminology,Craver:PubKeyStego}.

\subsubsection{Channel Noise}

In general, all covert channels can be \emph{noisy} since network frames or packets of the overt channels can be reordered, modified or lost, which can lead to bit errors or bit deletions/erasures in the covert channel. However, storage channels exploit the fact that overt protocols, such as TCP, have mechanisms for reliable data transport. If the header fields in which the covert channels are encoded are not changed in the network, these channels are effectively \emph{noise-free}. Timing channels on the other hand are always noisy, since the network always affects the timing of frames, packets or messages depending on the network conditions (e.g. congestion).

Active wardens (e.g. traffic normalizers) may also introduce noise. However, we do not consider this type of noise as part of the covert channel characteristics.

\subsubsection{Network Covert Channels}

The majority of the network covert channels proposed early were storage channels, for example Girling proposed embedding hidden information in address fields \cite{Girling87} and Rowland suggested embedding covert channels in different unused areas in the IPv4 header and in the TCP header \cite{Rowland97}. Storage channels are easy to use, as the header fields used are not modified inside the network and hence there is limited channel noise. However, most of these channels are also easy to detect and eliminate.

A few timing channels were proposed early on, for example Wolf presented a covert channel based on the timing of message acknowledgements \cite{WolfCCsLAN}. However, timing channels have only received more attention in recent years, for example Cabuk \emph{et al.} proposed to encode covert data into varying packet rates over time \cite{Cabuk06} and Berk \emph{et al.} proposed to encode covert bits by manipulating the gaps between consecutive packets (inter-packet gaps) \cite{berk05}. Timing channels are harder to use and always noisy, but they are also harder to detect and to eliminate.

Network covert channels can either be active by generating their own traffic or can piggyback on traffic created by a third party \cite{Rutkowska04chaoscommunication,JitterBug} to increase the channel's covertness.

\subsection{Related Work}

We now describe existing surveys and classifications of network covert channels, and how our novel pattern-based classification improves on these.

An early taxonomy of covert channels in multilevel security systems (MLS) was presented by Meadows and Moskowitz \cite{CC:ContextBasedView}. Covert channels were associated with four different contexts based on the service conditions in which they occur: High-to-low service covert channels, low-to-high service covert channels, shared service covert channels, and incomparable service covert channels. The taxonomy in \cite{CC:ContextBasedView} concentrates on early covert channel techniques that break security policies but do not necessarily provide stealthy communication. Our work focuses on network covert channels and on their hiding techniques instead of the service conditions in which they appear.

Shen \emph{et al.} classified local covert channels and proposed the idea to counter local covert channels based on their characteristics \cite{OptimizationofCCIdentification}. In comparison, our work discusses network covert channels and provides a hierarchical and more extensive categorization.

Llamas \emph{et al.} surveyed covert channels in Internet protocols \cite{Llamas:Survey}. The first part of the paper covers fundamentals of covert channel research for both, local and network covert channels. The second part summarizes publications on network covert storage and timing channels in TCP/IP protocols. However, \cite{Llamas:Survey} merely lists the different covert channels and makes no attempt to categorize them.

In 2007, Zander \emph{et al.} \cite{journals/comsur/ZanderAB07} published a comprehensive survey on network covert channels. The work covers the terminology, adversary scenario, covert channel techniques and countermeasures. Moreover, a categorization is applied which differs between channels taking advantage of unused header bits, header extensions and padding, the IP Identifier and the Fragment Offset, the TCP Initial Sequence Number (ISN), checksum fields, the Time to Live (TTL) field, the modulation of address fields and packet lengths, the modulation of timestamp fields, 
packet rate and timing, message sequence timing, packet loss and packet sorting, frame collisions, ad-hoc routing protocol-based techniques, 
Wireless LAN techniques, HTTP- and DNS-based techniques, application layer protocol-based channels and payload tunneling. The categorization 
in \cite{journals/comsur/ZanderAB07} is fine-grained: channels are not only categorized by their underlying technique, but also by the protocol layer they operate on. Also, \cite{journals/comsur/ZanderAB07} does not provide a hierarchy or a standardized pattern definition.

Zhiyong and Yong proposed a taxonomy based on entropy, and their work is the closest to our own \cite{EntropyTaxonomyNCC}. Each channel falls into one out of three categories depending on the `source' used to encode the covert data: variety entropy, constant entropy or fixed entropy. As in our own work, \cite{EntropyTaxonomyNCC} motivates the development of prevention techniques with a focus on covert channel categories instead of single techniques. We propose a hierarchical and more fine-grained categorization, which is not based on entropy but on the actual hiding techniques. While the development of \emph{more detailed and particular} countermeasures for the provided classification was left for future work in \cite{EntropyTaxonomyNCC}, our categorization enables more practical countermeasures that can address covert channel patterns.

We define an improved network covert channel taxonomy based on PLML/1.1 patterns and present the concepts of pattern variation, pattern combination and pattern hopping. Pattern \emph{variation} allows the adaption of one hiding technique to arbitrary network protocols. Pattern \emph{combination} allows to simultaneously use multiple patterns for a single PDU (e.g. a single network packet) or for a sequence of PDUs, while pattern \emph{hopping} randomizes multiple pattern-based covert channels to improve stealthiness. Pattern hopping enables a channel to adapt itself to changing conditions in networks (e.g. switching to another pattern if one pattern is blocked). Another advantage over the existing surveys is the fact that we include recent publications from 2009 to 2013 into our categorization.
Besides presenting the patterns, we also discuss countermeasures in the context of the identified patterns.

\section{Pattern-related Fundamentals and Their Taxonomy Use}\label{Sect:FundamRelWorkPatterns}

After introducing fundamentals and related work on covert channels, we now cover the fundamentals of patterns and pattern languages. Moreover,we describe our use of PLML.

\subsection{Patterns and Pattern Languages}
Graphical notations like UML 2.0 are powerful modeling languages ​​for the description of specifications and the subsequent documentation \cite{UML23} but only represent the end result of the design process.
Since the late 70s \emph{patterns}, in comparison, enable the successful documentation of design decisions during the development process.
In 1977, Alexander introduced first design patterns for solving problems in architecture and urban planning \cite{Alexander77}. Eighteen years later, the \emph{Gang of Four} (GoF) transferred the pattern concept to the domains of software architecture and software engineering \cite{GoF94}. Today, patterns are also used in fields of human computer interaction (HCI) research \cite{CHI2003}, usability engineering \cite{UE}, user experience \cite{UX}, task modeling \cite{Gaffar2004}, and application security \cite{Yoder98architecturalpatterns}. 

\begin{figure}[!b]
\centering
\includegraphics[width=0.75\textwidth]{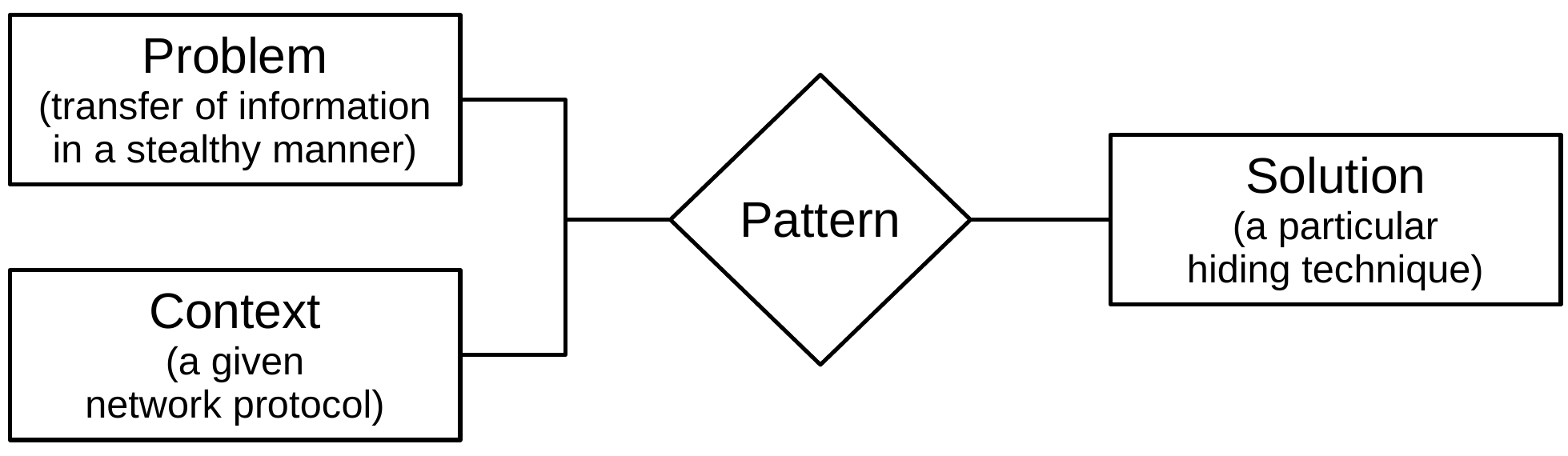}
\caption{The concept of patterns}
\label{fig:PatRel}
\end{figure}

As shown in Fig.~\ref{fig:PatRel}, all patterns represent a relation between a certain design problem and a solution in a given context. The \emph{problem} is a description of the issue to be solved. The \emph{solution} refers to a specific design that solves the given problem. The \emph{context} describes a repetitious set of cases in which the pattern can be used.

The use of patterns has a number of advantages \cite{Seffah10}: patterns are simple and easily readable for designers, developers and researchers, and they are useful for the collaboration between the people involved.
Furthermore, patterns are based on established knowledge and capture fundamental principles for good designs. Patterns also specify requirements in a general way that allows different implementations of the same pattern.


\subsection{Utilization of PLML for a Covert Channel-based Taxonomy}

In order to ensure a certain standard, patterns are summarized in a so-called \emph{pattern catalog} \cite{Alexander77}. A catalog of related patterns that belong to a common domain is a so-called \emph{pattern language} \cite{Seffah10}. Today, widely accepted pattern catalogs exist in the field of HCI, such as the catalogs presented by van Welie \cite{Welie13}, Tidwell \cite{tidwell09}, and van Duyne \cite{van07}. However, these pattern catalogs are described in different styles. 

To enable the clear definition and comparison of patterns, so-called \emph{schemes} were introduced \cite{Alexander77}. These schemes are divided into sections of textual and graphical descriptions. However, no standardized description of pattern scheme \emph{attributes} existed, as numerous pattern catalogs were created, based on a different understanding of attributes.
Due to these inconsistencies, searching and referencing patterns across different catalogs is difficult. To overcome this problem, a standardized pattern language was developed based on XML: the \emph{Pattern Language Markup Language} (PLML) v.\ 1.1 \cite{CHI2003}. PLML unifies and standardizes the schemes of different authors with the help of XML tags. Each XML tag represents a part of the scheme. Tab.~\ref{tab:PLML:Attrib} describes the tags from the PLML 1.1 we used to define covert channel patterns.

\begin{table}[!thb]

\tbl{Used PLML/1.1 Attributes\label{tab:PLML:Attrib}}
 { \begin{tabular}{ll}
   \toprule
   \textbf{Tag}			&	\textbf{Description} \\
   \midrule
	$<$pattern id$>$	&	Identifies a pattern within the particular catalog. \\
   \hline
	$<$name$>$ 		&	A correct assignment of a name for each pattern is important for the retrieval of a \\
				&	pattern when the pattern becomes part of a second catalog.\\

   \hline
   	$<$alias$>$ 		&	Patterns can have different names, which are specified in the $<$alias$>$ tag.\\
				&	The alias tag helps to find the same pattern when the pattern has different names\\
				&	in different catalogs.\\

   \hline
   	$<$illustration$>$ 	&	An application scenario for the pattern.\\

   \hline
   	$<$context$>$		&	Specifies the situations to which the pattern can be applied.\\

   \hline
   	$<$solution$>$		&	Describes the solution for a problem to which the pattern can be applied.\\
				&	The attributes \emph{problem} and \emph{context} (cf.\ Fig.~\ref{fig:PatRel}) are usually blurred but often not \\
				&	separated into two attributes.\\

   \hline
   	$<$evidence$>$		&	Contains additional details about the pattern and its design. Moreover, the tag can \\
				&	contain examples for known uses of the pattern. 
					\\
					
   \hline
   	$<$literature$>$	&	Lists references to publications related to the pattern.\\

   \hline
   	$<$implementation$>$	&	Introduces existing implementations, code fragments or implementational.\\
   \bottomrule
  \end{tabular}
  }

\end{table}

\section{Classification of Covert Channel Patterns by using PLML}\label{Sect:CCPatterns}

We evaluated the covert channel research of the last decades and classified the \NumOfAnalyzedTechniques{} evaluated covert channel techniques into \NumOfPatterns{} abstract patterns.

\subsection{Coverage of Techniques}

To select the most significant covert channel techniques for our pattern catalog, we took the existing surveys by Llamas \emph{et al.}~\cite{Llamas:Survey}, Zander \emph{et al.}~\cite{journals/comsur/ZanderAB07} and Zhiyong and Yong~\cite{EntropyTaxonomyNCC} into account and included the referenced publications in our evaluation.  Moreover, we cover additional papers with a significant amount of citations or novelty that were not mentioned in the surveys (e.g. because of their later publication between 2009 and 2013).

\subsection{Pattern List}

We now describe all patterns. 
For better readability, in the following textual presentation of the pattern catalog we merged the content of the \emph{literature} and \emph{evidence} attributes in the \emph{evidence} attribute and removed pattern attributes from PLML which are either redundant or not pertinent to our contribution.
Some papers propose various techniques belonging to the same pattern
(e.g. \cite{springerlink:10.1007/11767831_10} presents 10 hiding
techniques forming part of the \textit{Add Redundancy} pattern). In such cases, we
do not mention all techniques explicitly.

For some patterns less than three use cases exist in the literature. In
such cases, we added our own ideas for hiding techniques to provide
a minimum of three use cases. Our hiding techniques have no citation, as they are
initially proposed in this paper.

\newcommand{\PatEHead}{\vspace{0.2cm}\noindent}
\newcommand{\PatEHeadEnd}{\vspace{0.2cm}}
\newcommand{\PatE}{\noindent \hangafter=1 \hangindent+20pt}

\pagebreak
\PatEHead \textbf{P1. Size Modulation Pattern:}\PatEHeadEnd

   \PatE  \emph{Illustration:} The covert channel uses the size of a header element or of a PDU to encode the hidden message.

    \PatE \emph{Context:} Network Covert Storage Channels $\rightarrow$ Modification of Non-Payload $\rightarrow$ 
			 	Structure Modifying








    \PatE \emph{Evidence:}\\
    		1. Modulation of data block length in LAN frames \cite{Girling87}\\ 
		2. Modulation of padding field's size in IEEE 802.3 frames \cite{WolfCCsLAN} \\ 
		3. Modulation of IP fragment sizes \cite{MurdochLewis05,IPFragmentation}\\
		4. Modulate the message length of network packets \cite{NormalTrafficNCC}\\
		5. Modulate the size of IPSec messages \cite{SilenceInLANs}\\
		6. A man-in-the-middle adversary between VPN sites actively manipulates the maximum transmission unit (MTU) within the \emph{path MTU discovery} process between the VPN sites. Path MTU discovery is a continous process and changed MTUs are propagated to systems within the VPN site, i.e. allow to encode hidden information within the MTU \cite{SilenceInLANs}.

\PatEHead \textbf{P2. Sequence Pattern:}\PatEHeadEnd

   \PatE  \emph{Illustration:} The covert channel alters the sequence of header/PDU elements to encode hidden information.

   \PatE \emph{Context:} Network Covert Storage Channels $\rightarrow$ Modification of Non-Payload $\rightarrow$ 
			 	Structure Modifying

   \PatE  \emph{Evidence:}\\
			1. Sequence of Hypertext Transfer Protocol (HTTP) header fields \cite{DyatlovCastro05} \\
            2. Sequence of Dynamic Host Configuration Protocol (DHCP) options \cite{DHCP:CovChan}\\
            3. Sequence of File Transfer Protocol (FTP) commands \cite{FtpCmdSeqCC}

\PatEHead \textbf{P2.a. Position Pattern:}\PatEHeadEnd

   \PatE \emph{Illustration:} The covert channel alters the position of a given header/PDU element to encode hidden information.

   \PatE \emph{Context:} Network Covert Storage Channels $\rightarrow$ Modification of Non-Payload $\rightarrow$ 
			 	Structure Modifying $\rightarrow$ Sequence
               
   \PatE \emph{Evidence:}\\
   			1. Position of an IPv4 option in the options list of an IPv4 packet\\
            2. Position of an IPv6 extension header in the list of extension headers\\
			3. Position of a DHCP option in the options list \cite{DHCP:CovChan}

\PatEHead \textbf{P2.b. Number of Elements  Pattern:}\PatEHeadEnd

    \PatE \emph{Illustration:} The covert channel encodes hidden information by the number of header/PDU elements transferred.

    \PatE \emph{Context:} Network Covert Storage Channels $\rightarrow$ Modification of Non-Payload $\rightarrow$ 
			 	Structure Modifying $\rightarrow$ Sequence 

    \PatE \emph{Evidence:}\\
    			1. Alter the number of options placed in an IPv4 packet\\
                2. Modulate the number of options placed in a DHCP packet \cite{DHCP:CovChan}\\
			 	3. Modulate the number of fragments created from an original IP packet \cite{IPFragmentation}

\PatEHead \textbf{P3. Add Redundancy Pattern:}\PatEHeadEnd

    \PatE \emph{Illustration:} The covert channel creates new space within a given header element or within a PDU to hide data into.

    \PatE \emph{Context:} Network Covert Storage Channels $\rightarrow$ Modification of Non-Payload $\rightarrow$ 
			 	Structure Modifying 

    \PatE \emph{Evidence:}\\
    		1. Generation of packets with IPv4 options that embed hidden data \cite{Trabelsi10}\\
            2. Create a new IPv6 destination option with embedded hidden data \cite{Graf03} \\
            3. Extend HTTP headers with additional fields or extend values of existing fields \cite{DyatlovCastro05} \\
            4. Manipulate the \emph{pointer} and \emph{length} values for the IPv4 record route option to create space for data hiding \cite{Trabelsi10} \\
            5. Add random bytes to an encrypted SSH message \cite{lucena04}\\
            6. Extend Simple Mail Transfer Protocol (SMTP) packet headers with additional fields \cite{SMTPCCs}\\
            7. Hide data in unused bits of the DHCP \emph{chaddr} field if the \emph{hlen} field is set to a value that is larger than the size of a network address \cite{DHCP:CovChan}\\
            8. Encapsulate IP packets with a smaller size than specified in the Ethernet frame size and use the space between the end of the IP packet and the Ethernet trailer for covert data \cite{muchene2013reporting}\\
            9. Encode hidden information through the presence/absence of ``type'' or ``xml:lang'' attributes in the Extensible Messaging and Presence Protocol (XMPP) or through the presence of leading/trailing white spaces in XMPP messages \cite{CCsInXMPP}

\PatEHead \textbf{P4. PDU Corruption/Loss Pattern:}\PatEHeadEnd

    \PatE \emph{Illustration:} The covert channel generates corrupted PDUs that contain hidden data or actively utilizes packet loss to signal hidden information.

    \PatE \emph{Context:} Network Covert Storage Channels $\rightarrow$ Modification of Non-Payload $\rightarrow$ 
			 	Structure Modifying 

    \PatE \emph{Evidence:}\\
    			1. Generate corrupted messages in broadcast erasure channels \cite{Phantomes}\\ 
			2. Transfer corrupted frames in IEEE 802.11 \cite{WLANStego}\\
 	                3. A man-in-the-middle adversary between two VPN sites drops selected packets exchanged between
 	                	the VPN sites to introduce covert information into an established connection of
 	                	adversaries located within the VPN sites \cite{SilenceInLANs}.
     

\PatEHead \textbf{P5. Random Value  Pattern:}\PatEHeadEnd

    \PatE \emph{Illustration:}  The covert channel embeds hidden data in a header element containing a ``random'' value.

    \PatE \emph{Context:} Network Covert Storage Channels $\rightarrow$ Modification of Non-Payload $\rightarrow$ 
		 	Structure Preserving $\rightarrow$ Modification of an Attribute

    \PatE \emph{Evidence:}\\
    		1. Utilize the IPv4 \emph{Identifier} field \cite{Rowland97}\\
    		2. Utilize the first sequence number of a TCP connection -- the Inital Sequence Number (\emph{ISN}) \cite{Rowland97,Rutkowska04chaoscommunication}\\
            3. Hide data in the TCP ISN using a bounce server \cite{Rowland97}\\
            4. Utilize the DHCP \emph{xid} field \cite{DHCP:CovChan} \\
            5. Utilize the Secure Shell (SSH) protocol Message Authentication Code (\emph{MAC} field) \cite{lucena04}

    
    \PatE \emph{Notes:} As some header elements, such as the TCP ISN, follow a 
			    distribution which conforms to a particular operating system 
             	or context, their values cannot be considered perfectly random 
             	and the placement of ``random'' values in such elements can lead to 
                different value distributions, which can be detected \cite{MurdochPhDThesis}.

\PatEHead \textbf{P6. Value Modulation Pattern:}\PatEHeadEnd

    \PatE \emph{Illustration:} The covert channel selects one of $n$ values a header element can contain to encode a hidden message.

    \PatE \emph{Context:} Network Covert Storage Channels $\rightarrow$ Modification of Non-Payload $\rightarrow$ 
		 	Structure Preserving $\rightarrow$ Modification of an Attribute

    \PatE \emph{Evidence:}\\
    		1. Send a frame to one of $n$ available Ethernet addresses in the local network \cite{Girling87}\\ 
		 	2. Encode information by $n$ of the possible IP header Time-to-live (TTL) values (e.g. a high 
		 	or a low TTL value) \cite{ZanderEtAl:TTL:Capacity} \\
            3. Encode information by $n$ of the possible \emph{Hop Limit} values in the IPv6 header \cite{springerlink:10.1007/11767831_10}\\
            4. Encode information by sending a 
		 	packet using one of $n$ possible application layer protocols \cite{wendzel12detecting} or application layer ports \cite{BorlandBlog}\\
            5. Encode information by selecting one of $n$ possible messages types in the Building Automation and Control Networking (BACnet) protocol  \cite{WendzelBACnet}\\
            6. Encode information in the target IP of address resolution protocol (ARP) messages \cite{CovChanLAN}\\
            7. Change the value of the ``type'' or ``xml:lang'' attributes in XMPP \cite{CCsInXMPP}\\
            8. Send IPSec packets from one VPN site to specific destination IPs within another VPN site \cite{SilenceInLANs}.

\vspace{0.15cm}

\PatEHead \textbf{P6.a. Case Pattern:}\PatEHeadEnd

    \PatE \emph{Illustration:} The covert channel uses case-modification of letters in header elements to encode hidden data. 

    \PatE \emph{Context:} Network Covert Storage Channels $\rightarrow$ Modification of Non-Payload $\rightarrow$ 
		 	Structure Preserving $\rightarrow$ Modification of an Attribute $\rightarrow$ 
		 	Value Modulation 

    \PatE \emph{Evidence:}\\
    		1. Case modification in HTTP headers \cite{DyatlovCastro05}\\
    		2. Modify the case of the ``type'' or ``id'' attributes in XMPP \cite{CCsInXMPP}\\
    	    3. Case modification in SMTP, Post Office Protocol (POP3), or Network News Transfer Protocol (NNTP), commands and headers

\vspace{0.15cm}
     
\PatEHead \textbf{P6.b. Least Significant Bit (LSB) Pattern:}\PatEHeadEnd

    \PatE \emph{Illustration:} The covert channel uses the least significant bit(s) of header elements to encode hidden data. 

    \PatE \emph{Context:} Network Covert Storage Channels $\rightarrow$ Modification of Non-Payload $\rightarrow$ 
		 	Structure Preserving $\rightarrow$ Modification of an Attribute $\rightarrow$ 
		 	Value Modulation 

    \PatE \emph{Evidence:}\\
            1. Encode into the IPv4 timestamp option by effectively sending at even/odd times \cite{HidingDataintheOSINetworkModel}\\
            2. Modify the low order bits of the timestamp option in TCP \cite{Giffin:2002}\\
            3. Utilize the least significant bits of the \emph{secs} field in the DHCP header \cite{DHCP:CovChan}\\
            4. Encode covert bits in slight modifications of view angles (yaw, pitch) of player's avatars in the Quake3 multiplayer game protocol \cite{Zander:MPGameCCs} \\
            5. Utilize the least significant bits of the IPv4 \emph{TTL} field\\ 
	    6. Utilize the least significant bits of the IPv6 \emph{Hop Limit} field \cite{springerlink:10.1007/11767831_10}\\
            7. Utilize the least significant bits of the \emph{Hop Count} field in the network layer PDU of the BACnet protocol\\
            8. Utilize the least significant bits of the ``id'' attribute in XMPP \cite{CCsInXMPP}

\vspace{0.15cm}
\PatEHead \textbf{P7. Reserved/Unused Pattern:}\PatEHeadEnd

    \PatE \emph{Illustration:} The covert channel encoded hidden data into a reserved or unused header/PDU element.

    \PatE \emph{Context:} Network Covert Storage Channels $\rightarrow$ Modification of Non-Payload $\rightarrow$ 
		 	Structure Preserving $\rightarrow$ Modification of an Attribute

    \PatE \emph{Evidence:}\\
    		1. Utilize undefined/reserved bits in IEEE 802.5/data link layer frames \cite{WolfCCsLAN,HidingDataintheOSINetworkModel}\\
            2. Utilize unused fields in IPv4, e.g. Identifier field, Don't Fragment (DF) flag or the reserved flag, as well as in IP-IP encapsulation \cite{HidingDataintheOSINetworkModel,Ahsan,buchanan2004covert,SilenceInLANs}\\
            3. Encode hidden data in unused or reserved fields of the IPv6 header or its extension headers (\cite{springerlink:10.1007/11767831_10} lists 8 hiding techniques for the Reserved/Unused pattern in IPv6) \\
            4. Utilize unused bits in the TCP header \cite{HidingDataintheOSINetworkModel}\\
            5. Utilize the ICMP echo payload \cite{pingtunnel,LOKI2}\\
            6. Utilize the padding field of IEEE 802.3 \cite{WolfCCsLAN,ImprFramePad}\\
            7. Utilize unused fields in the BACnet header \cite{WendzelBACnet}\\
            8. Place hidden data behind the string termination symbol in the \emph{sname} and \emph{file} fields of DHCP \cite{DHCP:CovChan}\\
            9. Place hidden information into the \emph{Differentiated Services} (DS) field of outbound IPSec connections \cite{SilenceInLANs}.\\
            10. Insert hidden data into the IP \emph{Explicit Congestion Notification} (ECN) field in IPSec connections \cite{SilenceInLANs}.

\PatEHead \textbf{P8. Inter-arrival Time Pattern:}\PatEHeadEnd

    \PatE \emph{Illustration:} The covert channel alters timing intervals between network PDUs (inter-arrival times) to encode hidden data.

    \PatE \emph{Context:} Network Covert Timing Channels 

    \PatE \emph{Evidence:}\\
    		1. Alter the timings between LAN frames sent \cite{Girling87}\\
            2. Alter the response time of a HTTP server \cite{EsserRWTHApacheTimingC}\\
            3. Alter the timings between BACnet/IP packets \cite{WendzelBACnet}\\
            4. Introduce artificial delays into inter-arrival times of SSH packets sent based on keyboard input (interactive shell) \cite{JitterBug}\\
            5. Acknowledge IEEE 802.2 I-format frames immediately or after a second I-format frame was received \cite{WolfCCsLAN}\\
            6. A man-in-the-middle (MitM) adversary in the public network between
		           two VPN-secured sites modifies the inter-arrival times of packets
		           transferred between two man-in-the-edge (MitE) systems on each site of
		           the VPN to signal hidden data to both MitE adversaries \cite{ARES:MiTM:MitE:CCs}.\\
	          7. Alternatively to (6), the MitE systems communicate covertly by sending traffic with manipulated inter-arrival times to each other \cite{SilenceInLANs,ARES:MiTM:MitE:CCs}.\\
	       	8. Record a legitimate traffic sequence, partition the sequence and replay the inter-arrival times of a particular partition \cite{Cabuk06}

\PatEHead \textbf{P9. Rate Pattern:}\PatEHeadEnd

	\PatE \emph{Illustration:} The covert channel sender alters the data rate of a traffic flow from 
		 	itself or a third party to the covert channel receiver.

	\PatE \emph{Alias:} Throughput Pattern

    \PatE \emph{Context:} Network Covert Timing Channels 

    \PatE \emph{Evidence:}\\
    		1. Exhaust the performance of a switch to affect the throughput 
		 	of a connection from a third party to a covert channel 
		 	receiver over time \cite{SwitchCovertChan}\\
            2. Manipulate the serial communication port's throughput by 
		 	delaying \emph{Clear to Send}/\emph{Ready to Send} commands \cite{HidingDataintheOSINetworkModel}\\
		 	3. Directly alter the data rate of a legitimate channel between a 
		 	covert channel sender and receiver. 

\PatEHead \textbf{P10. PDU Order Pattern:}\PatEHeadEnd

    \PatE \emph{Illustration:} The covert channel encodes data using a synthetic PDU order for a 
				 	given number of PDUs flowing between covert sender and receiver. 

    \PatE \emph{Context:} Network Covert Timing Channels 

    \PatE \emph{Evidence:}\\
    		1. Modify the order of IPSec Authentication header (AH) packets \cite{Ahsan}\\
            2. Modify the order of IPSec Encapsulated Security Payload (ESP) packets \cite{Ahsan}\\
            3. Modify the order of TCP packets \cite{Cloak,PktReordPh}.\\
            4. A MitM adversary in the public network between two VPN-secured sites modifies the order of packets transferred
		           between two MitE systems on each site of the VPN to signal hidden data to both MitE adversaries \cite{ARES:MiTM:MitE:CCs}.\\
	          5. Like (4), modify the order of IPSec packets for inbound or outbound VPN traffic \cite{SilenceInLANs}.\\
 	                6. A covert channel sender transfers frames in a way they are sent before or after a legitimate user's frames in CSMA/CD networks. The covert channel receiver analyzes the order of arriving frames \cite{HidingDataintheOSINetworkModel}.\\


\PatEHead \textbf{P11. Re-Transmission Pattern:}\PatEHeadEnd

    \PatE \emph{Illustration:} A covert channel re-transmits previously sent or received PDUs.

    \PatE \emph{Context:} Network Covert Timing Channels 

    \PatE \emph{Evidence:}\\
		1. Transfer selected DNS requests once/twice to encode a hidden bit per request.\\
                2. Duplicate selected IEEE 802.11 packets \cite{WLANStego}\\
       		3. Encode hidden data by re-transmitting selected TCP segments.\\
       		4. Do not acknowledge received packets in order to force the covert sender to re-transmit a packet. The re-transmitted packet is modified by the sender to carry hidden data \cite{Mazurczyk:ReTransmission}.

\subsection{Taxonomy/Classification}

\begin{figure*}[!thb]
\centering
\includegraphics[width=1.00\textwidth]{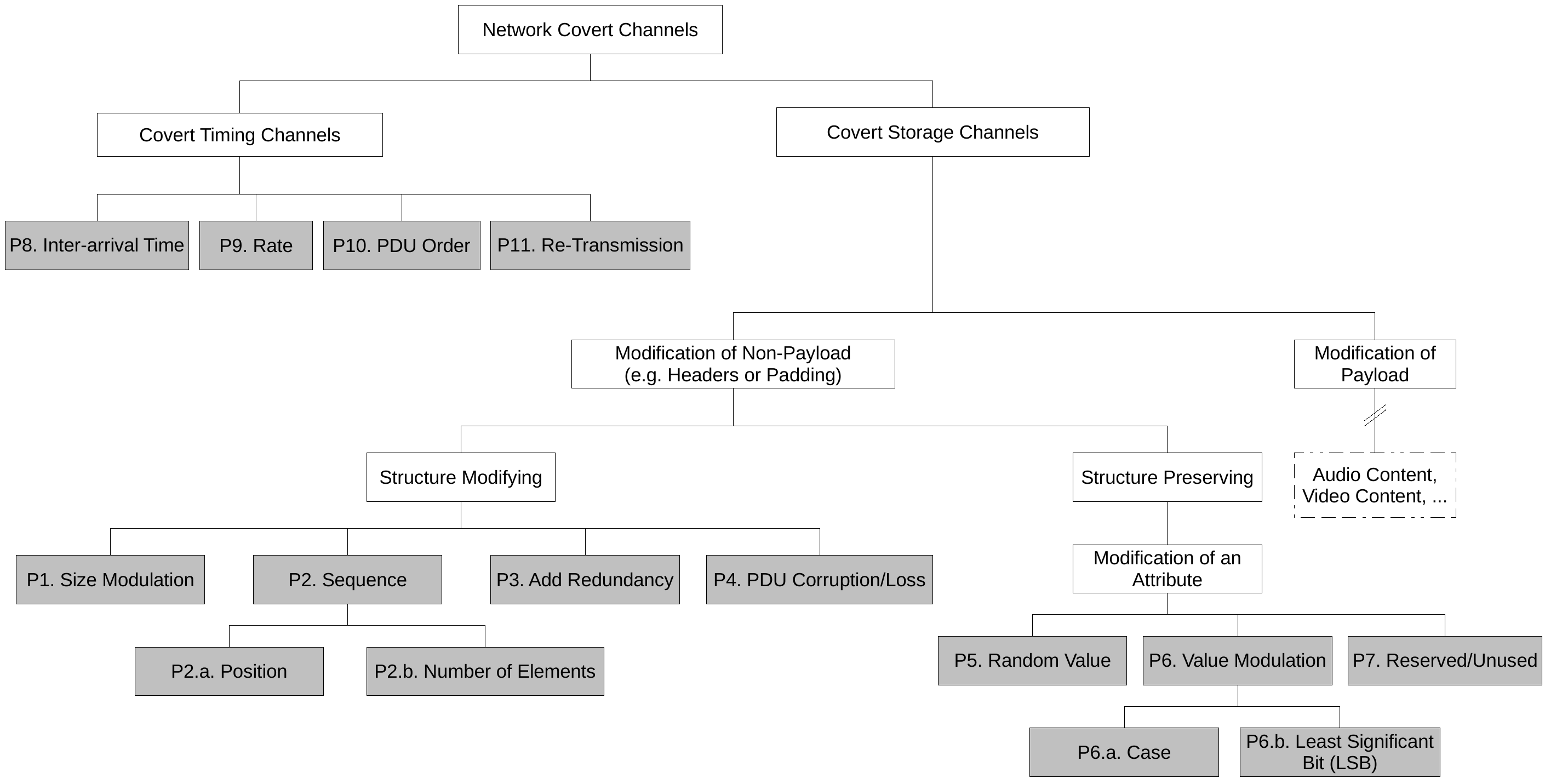}
\caption{Network covert channel pattern hierarchy, excluding hiding techniques utilizing payload}
\label{fig:PatternHierarchy}
\end{figure*}

We provide a hierarchical view of the discovered patterns in order to structure our findings. The hierarchy is visualized in Fig.~\ref{fig:PatternHierarchy} where white boxes represent categories of patterns and gray boxes represent patterns. Conforming to the PLML standard, a covert channel pattern can also be a \emph{child pattern} of a parent pattern, as in case of the \emph{Case pattern}. 

The major categorization of all network covert channels is into timing and storage channels. We introduce additional sub-categories for storage channels due to their diversity. We distinguish between storage channels which apply hiding methods to payload (e.g. to audio streaming) -- these channels are outside of our scope -- and storage channels which alter non-payload (e.g. header elements or padding bits). These non-payload modifying channels do either change or preserve the structure of a PDU -- a novel difference we discovered in the analysis process.

In case of a structure modification, a pattern either alters the order of elements in the protocol header or it changes the size of the PDU. We discovered different patterns for both variants. On the other hand, if a pattern preserves the structure of a PDU, the pattern must modify a data element in the PDU (e.g. a header field).

Besides the given hierarchical representation, Tab.~\ref{tab:AdditionalPatternCategorization} categorizes all patterns regarding the following additional aspects:

\begin{itemize}
  \item \textbf{Semantic:} The semantic of a PDU is changed if the pattern modifies header elements in a way that leads to a different interpretation of the PDU. For instance, the semantic of an IPv4 header is changed if a ``record route'' option is attached but preserved if the reserved flag is set as the reserved flag does not lead to a changed interpretation of the packet.
  In general, a channel raises less attention if the semantic of network data is not modified.
    
  \item \textbf{Syntax:} We call a modification of the PDU structure a syntax modification. For instance, adding additional header elements changes the PDU structure.
  As with the semantic, a covert channel pattern can either modify or preserve the syntax. The syntax categorization is only applied to storage channels as timing channels do not change the structure of a PDU. 
   
   \item \textbf{Noise:} 
   In general, noise in the form of bit corruptions or packet loss can affect all presented patterns. Here we only categorize a pattern as noisy if the channel is a timing channel and thus, always faces noise, or if it is embedded in PDU fields which are modified in the network (like the TTL in the IPv4 header). Although active wardens can introduce additional noise in many patterns (e.g. by removing IPv4 options used to carry hidden data), we do not take normalization effects into account.
   
\end{itemize}

\begin{table}[!hb]

\tbl{Categorization of Covert Channel Patterns\label{tab:AdditionalPatternCategorization}}
 { \begin{tabular}{lcccccc}
   \toprule
         	    &\multicolumn{2}{c}{\textbf{Semantic}}& \multicolumn{2}{c}{\textbf{Syntax (Structure)}}	&	\multicolumn{2}{c}{\textbf{Noise}}	\\
   \textbf{}   &	preserving 	& modifying    	& preserving 		& modifying  		&	noisy 	& noiseless  	\\
   \midrule
   \textbf{Storage Channel Patterns}  \\
   \hline
    P1. Size
    Modulation		&	X			&	 			&	 				&	X				&	 		&	~~X$^b$	\\ 
    P2. Sequence	&	~~X$^a$		&	 			&	 				&	X				&	 		&	X 	  \\
    P2.a. Position	&	~~X$^a$		&	 			&	 				&	X				&	 		&	X		  \\
    P2.b. Number of
    Elements		&	X			&	 			&	 				&	X				&	 		&	X		  \\
    P3. Add
    Redundancy		&	X			&	  			&	 				&	X				&	 		&	X		  \\
    P4. PDU
    Corruption/Loss	&	~~-$^e$		&	~~-$^e$		&	 				&	X				&	X		&	 		\\

    P5. Random
    Value			&	X			&	 			&	X				&	 				&	 		&	X		  \\
    P6. Value
    Modulation		&	 			&	X			&	X				&	 				&	~~X$^f$	&	~~X$^f$	  \\
    P6.a. Case			&	X			&	 			&	X				&	 				&	 		&	X		  \\
    P6.b. LSB				&	 			&	X			&	X				&	 				&	X		&	 		  \\
    P7. Reserved/Unused
    				&	~~X$^c$		&	~~X$^c$		&	X				&	 				&	 		&	X		  \\
   \midrule
   \textbf{Timing Channel Patterns} \\
   \hline
    P8. Inter-arrival
    Time			&	X			&	 			&	-				&	-				&	X		&	 		\\
    P9. Rate			&	X			&	 			&	-				&	-				&	X		&	 		\\
    P10. PDU Order		&	~~X$^d$		&	~~X$^d$		&	-				&	-				&	X		&	 		\\
    P11. Re-Transmission
					&	X			&	 			&	-				&	-				&	X		&	 		\\
   \bottomrule
  \end{tabular}
  }
  \begin{tabnote}%
\tabnoteentry{$^a$}{The semantic of Sequence and Position patterns is only preserved if an utilized element's position or the sequence of elements have no effect on the PDU's semantic.}
\vskip2pt
\tabnoteentry{$^b$}{Fragmentation can cause noise for channels using the Size Modulation pattern since routers can fragment large packets into multiple smaller packets.}
\vskip2pt
\tabnoteentry{$^c$}{The semantic of a PDU \emph{can} change if the covert channel modifies currently unused elements (e.g. a set DF flag in the IPv4 header would prevent fragmentation). On the other hand, a utilization of currently unused elements can \emph{preserve} the semantic, e.g. if the covert channel sets the MF flag in IPv4 to zero, the modification of the Fragment Offset will not lead to a changed semantic.}
\vskip2pt
\tabnoteentry{$^d$}{If the channel utilizes a protocol that provides packet sorting at the receiver side, the PDU Order pattern can preserve the semantic of the data transfer, otherwise it can change the semantic.}
\vskip2pt
\tabnoteentry{$^e$}{Intentionally corrupted PDUs are not interpreted and thus do neither change nor preserve the semantic of a PDU.}
\vskip2pt
\tabnoteentry{$^f$}{A value modulation can lead to noise (e.g. if the IP TTL is used) but can also be noiseless (e.g. if the source address is modulated).}
\end{tabnote}%

\end{table}

\subsection{Occurrence Rate of Particular Patterns}

\begin{figure}[!t]
 \centering
 \includegraphics[width=0.8\textwidth]{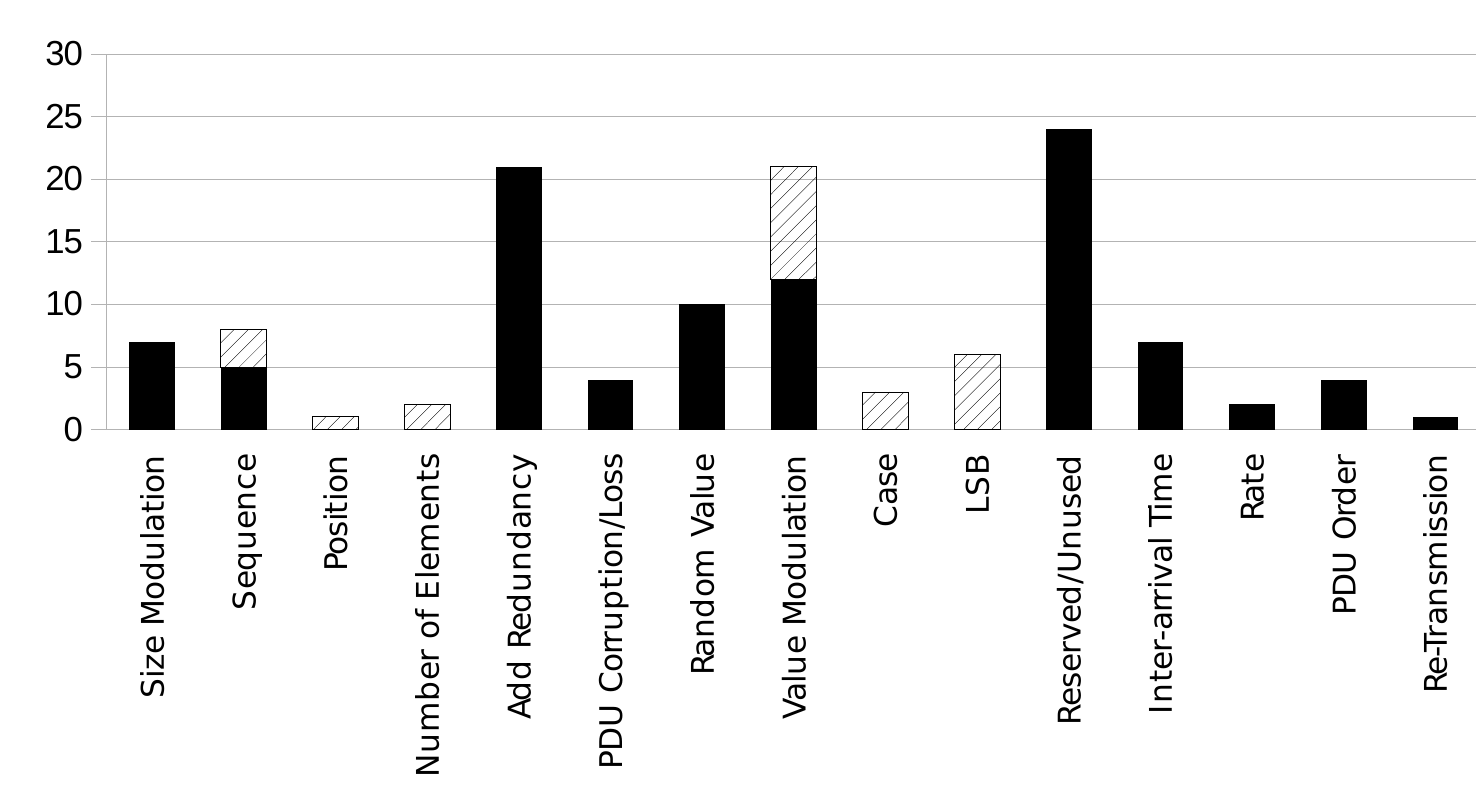}
 \caption{Number of associated covert channel techniques per covert channel pattern. Shaded bars represent child patterns.}
 \label{fig:PatAnzahl}
\end{figure}

Besides the fact that we were able to `reduce' the \NumOfAnalyzedTechniques{} hiding techniques to only \NumOfPatterns{} patterns, our findings also show that the majority (76 of \NumOfAnalyzedTechniques{} or \ProzentOfPatternsInTopCategories\%) of hiding techniques is based on only four patterns, namely the Reserved/Unused pattern (24 techniques), the Add Redundancy pattern (21 techniques), the Value Modulation pattern (21 techniques, including its child patterns Case and LSB), and the Random Value pattern (10 techniques). In other words, many of the surveyed and analyzed techniques are of relatively little novelty as they are based on existing techniques. Fig.~\ref{fig:PatAnzahl} compares the number of covert channel techniques associated with the particular patterns.

It would be interesting to compare the occurrence rate of patters in the literature with their actual number of uses in practice. However, no information about the usage rates of covert channels are available.

\subsection{Extensibility of the Pattern Catalog}

The patterns introduced in our work will be made available publicly online in a moderated wiki. A wiki allows the collaboration among experts from both, research institutions and industry: it allows active discussion and, through moderation, the controlled extension and modification of the pattern catalog.

If a novel hiding technique requires the integration of a new pattern, researchers can add the pattern to the catalog, which will be invisible until accepted by the moderators. The acceptance may be performed after detailed discussion with the contributing researchers to ensure the wiki's accurate and consistent state with the existing pattern collection and to prevent future contributors from wrongly classifying their ``new'' covert channel techniques as new patterns. Moreover, researchers can add upcoming hiding techniques that are based on an already existing pattern to the \emph{evidence} property of the particular pattern.

\section{Variation of Covert Channel Patterns}\label{Sect:Transformation}

Covert channels are not only used for malicious purposes (e.g. the control of botnets) but are also used to support the freedom of speech (e.g. of journalists). Furthermore, research on covert channel improvements enables the further development of necessary countermeasures against malicious users; the research community must identify improvements for covert channel techniques before malicious users take advantage of them.

We will now describe the variation of the patterns defined in the previous section. A pattern variation automatically adapts a pattern to a new context. In case of network covert channels, we define the \emph{utilized network protocol}, used as carrier for the hidden data, as the \emph{context}. If the channel switches the protocol, the context changes as well and thus, the pattern must be adapted to the new context. For instance, a pattern that was previously applied to hide data in IPv4 can be adjusted to hide data in IPv6.

A pattern variation is only useful if it is an automatic process for both, the sender and the receiver. The automatic process generates new code that implements the pattern for the altered context. Thus, a pattern variation eliminates the necessity of programming a pattern from scratch if it needs to be adapted to another network protocol.

\begin{figure}[!t]
\centering
 \includegraphics[width=0.65\textwidth]{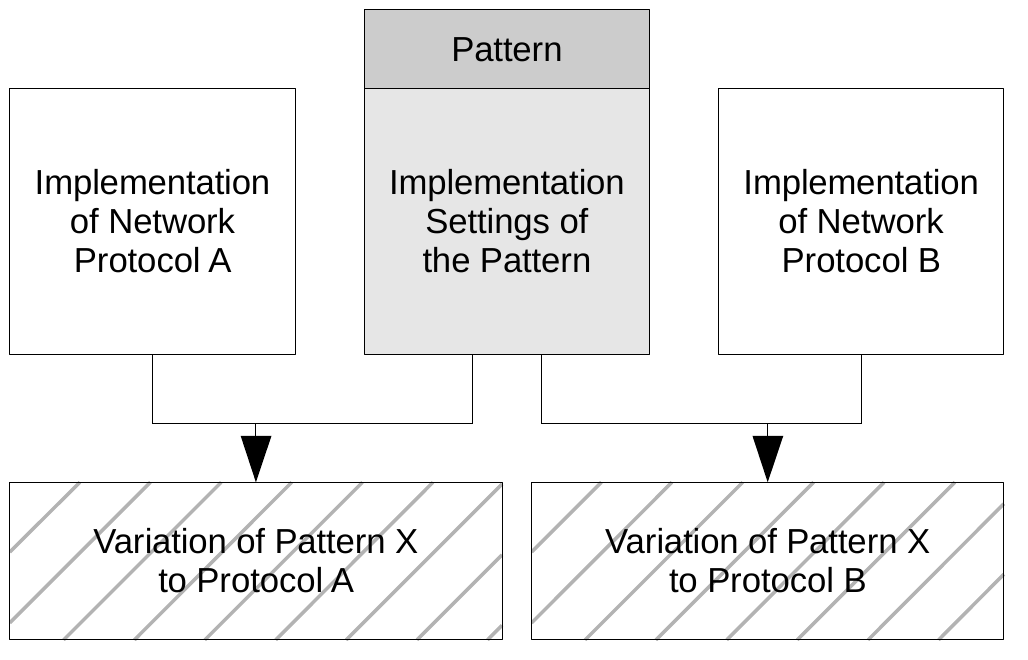}
 \caption{Concept of Covert Channel Pattern Variation}
 \label{fig:PatTransCC}
\end{figure}

Fig.~\ref{fig:PatTransCC} visualizes the concept of a covert channel pattern variation: While a general network protocol implementation is required to hide data within a protocol or the protocol's timing behavior, a pattern has information which enable the application of the pattern to other network protocols.

\subsection{PLML-based Pattern Variation}

Our approach is similar to that of Engel \emph{et al.} who store the information required for a pattern transformation within the \emph{implementation} property of PLML \cite{PatternTransformation,Engel:PatternTransNEU}. We  introduce so called \emph{settings} into the implementation property. These settings contain all significant information for a variation.

Engel \emph{et al.}\ define a configuration for each context of a pattern \cite{Engel:PatternTransNEU}. We match this aspect as well by introducing specific settings for each supported overt network protocol. For instance, a protocol $A$ may provide 4 bits of space for a pattern and the location of the bits is at an offset of 6 bits from the first bit of the header. In the new context of protocol $B$, the pattern can only use 3 bits and these 3 bits are located at an offset of 20 bits from the first bit of protocol $B$'s header. In other words, each variation of a pattern highly depends on the utilized network protocol. Given a very simple hiding technique that just utilizes a single bit and sets it to ``1'' or ``0'' in an arbitrary protocol with a static header structure, the variation would just need an ``offset'' value for each protocol to locate the selected bit in its header.

Figure~\ref{lst:SettingsExample} shows example settings for the \emph{Random Value} pattern. For IPv4, these settings would utilize the 16 bit Identifier field and for TCP, the 32 bit ISN would be used. As the ISN is only random in the first packet of a flow, a limitation (\emph{OnlyFirstPkt}) must be enforced.

\begin{figure}
\begin{lstlisting}
settings.ipv4.Offset=32;
settings.ipv4.Len=16;
settings.tcp.Offset=32
settings.tcp.Len=32
settings.tcp.OnlyFirstPkt=true
\end{lstlisting}
\caption{Settings for IPv4 and TCP in case of the Random Value Pattern}
\label{lst:SettingsExample}
\end{figure}

We found that the following settings can be necessary for covert channel patterns -- depending on the particular pattern:

\begin{itemize}
  \item \emph{Offset}: Number of bits between the first bit of the protocol header and the first bit of the utilized area
  \item \emph{Len}: Size of the utilized area
  \item \emph{OnlyFirstPkt}: Only use the first packet of a flow (e.g. for the TCP ISN)
  \item \emph{Min/MaxSize}: Minimum/maximum size of a (padding) field or of a frame
  \item \emph{Min/MaxElements}: Minimum/maximum number of elements to use (e.g. minimum number of IPv4 options or DHCP options for the Position pattern or the Sequence pattern)
  \item \emph{ValueRange/ValuesAllowed}: These fields limit the range of allowed values for a field and the particular values which can be stored in the field. For instance, the Value Modulation pattern may only be allowed to place the values ``a''-``j'' in a field or only the values ``yes'', ``no'' and ``optional''.
  For passive covert channels, i.e. those piggybacking third-party traffic, constraints can be defined in order to allow only limited modulations. For instance, a constraint can allow only TTL modifications in a range of +/-1.
  \item \emph{Min/Max/DistributionIPG}: Minimum/maximum time difference between packets or definition of a value distribution for these time differences. While this attribute is generally useful, it is especially required to configure limits for the Inter-arrival Time pattern.
  \item \emph{Min/MaxRate}: Minimum/maximum packet rate. This attribute is similar to the Min/MaxIPG attribute but configures the number of PDUs the channel is allowed to transfer per time $t$. The attribute is of general use but especially required to configure the limits for the PDU Order pattern to ensure stealthiness. In future work rate profiles could be defined as well in order to match the actual traffic behavior as close as possible.
\end{itemize}

While the actual hiding technique of a pattern can be adapted automatically, protocols with header fields which depend on other header fields still require the additional use of libraries and tools like \emph{scapy}\footnote{http://www.secdev.org/projects/scapy/}. For instance, the calculation of the IPv4 \emph{Checksum} as well as the adjustment of the \emph{Internet header length} and \emph{total length} fields are not included in the pattern variation process and must be done additionally. We therefore decided to include \emph{scapy} commands into our settings as \emph{scapy} automatically calculates values for header fields the user does not not explicitly define. Thus, the use of \emph{scapy} eliminates the need to implement a variation's utilized network protocols. Therefore, additional settings must be defined in the form
settings.\textbf{protocol}.value=\textbf{bit value},scapystring=\textbf{command}.

Figure~\ref{lst:ImplementationScapySettings} shows a sample setup for assigning two bit values to two different protocols (IPv4 and IPv6) of the LSB pattern using \emph{scapy} strings. A high and a low value are assigned to TTL or Hop Limit in order to transfer a ``0'' (\emph{value=0}) or a ``1'' (\emph{value=1}) bit. In order to prevent a trivial detection of the channel, the values are randomized in a higher/lower range.

\begin{figure}
\begin{lstlisting}
settings.ipv4.value=0,scapystring=IP(ttl=100+RandInt()%50);
settings.ipv4.value=1,scapystring=IP(ttl=150+RandInt()%50);
settings.ipv6.value=0,scapystring=IPv6(hlim=100+RandInt()%50);
settings.ipv6.value=1,scapystring=IPv6(hlim=150+RandInt()%50);
\end{lstlisting}
\caption{Sample tuples using \emph{scapy} strings for the LSB pattern}
\label{lst:ImplementationScapySettings}
\end{figure}

Like our pattern catalog, the settings proposed for pattern variation also serve as a basis for future work, i.e. they can be extended by adding additional settings in the future.

\subsection{Requirements-based Pattern Variation}

Pattern variation is also applicable in the context of situational requirements. For instance, the transfer of a video stream over a covert channel requires a higher throughput than the transfer of a password within the same time.

Each overt channel (the utilized network protocol) provides a particular amount of space to carry hidden data and has a different potential to raise attention \cite{CMSPaper}. A pattern variation for a given requirement can thus also switch the overt channel used in order to provide a high throughput or a high covertness. If the pattern is required to transfer a video stream it may use an overt channel providing a high throughput while the pattern may use an overt channel with a low throughput and high covertness if only a few hidden bits must be transferred. Therefore, a pattern cannot only change the overt channel but can also adjust the number of manipulated bits in the overt channel.

\subsection{Similar Approaches to Pattern Variation}

Covert channels can utilize multiple hiding techniques simultaneously \cite{CMSPaper,DBLP:conf/prdc/YarochkinDLHK08}. For instance, one technique could modify the LSB of the IPv4 TTL while another technique modifies the IPv4 reserved flag.

This section puts \emph{patterns} in the context of such a simultaneous application of hiding techniques. To this end, we describe two approaches that utilize multiple patterns instead of multiple hiding techniques, namely \emph{pattern combination} and \emph{pattern hopping}.



\subsubsection{Pattern Combination}

To increase the throughput of a covert channel and its stealthiness, we can combine multiple patterns, e.g. by applying them in parallel to a single packet -- an idea already shown for single hiding techniques in the field of network steganography \cite{MultiLevelStego} -- or sequentially to subsequent packets. 
As an example, consider the parallel application of the  Random Value and Add Redundancy patterns: A covert channel sender could modify the Identifier in the IPv4 header as well as it attaches an IPv4 option used to carry additional hidden data.
A parallel application of a particular set of patterns to a single packet may not always be possible. In future work we will determine dependencies between patterns so that feasible pattern combinations can be identified.
Much simpler is the sequential application of different patterns to subsequent packets. For instance, for one packet the Value Modulation pattern could be used to modify an unused field and LSB value modulation could be used for another field for the following packet.

\subsubsection{Pattern Hopping}

Sequential pattern combination uses a simple linear combination, which could be easily detected. To make detection more difficult, we propose a simple mechanism based on the concepts of protocol hopping covert channels \cite{CMSPaper} and synchronized random number generators similar to \cite{gianvecchio08}.

Let $P$ be a set of patterns. The sender $S$ and receiver $R$ agree to use a certain cryptographically secure pseudo-random number generator (CSPRNG) with a certain seed value $V$. The agreement on CSPRNG and $V$ has to be done separately over a secure transmission channel, since the covert communication could be easily uncovered if both are known.
The sender $S$ and receiver $R$ initialize their CSPRNG with $V$. Both CSPRNGs now are synchronized. Let $t$ be the sequential number of the transferred pattern, incremented each time a packet is being sent (and received). $t$ is initialized with 0. For example, $t$ could also be the timestamp or the sequence number of a packet (immutable between $S$ and $R$). $S$ chooses $p_i \in P$ where $i=\textrm{CSPRNG}(V,t)\:\textrm{mod}\:|P|$ and applies $p_i$
to the actual packet to send. $R$ knows the pattern since $R$ gets $t$ and knows $V$ and CSPRNG. Thus, the patterns used are randomized. Instead of using a pattern for each packet we can also increase the modulus so that `unmapped' patterns are ignored (limiting the bandwidth).

As network packets, and with it, the transfer of covert data, can be error-prone due to packet loss or re-transmissions caused by transport layer protocols, a reliable communication is a necessity to prevent the de-synchronization of the CSPRNG. To overcome this synchronization problem, so-called \emph{micro protocols} can be used, i.e. covert channel-internal control protocols with reliability features \cite{DBLP:conf/pst/RayM08}.  Micro protocols have been well studied over the last years and optimized micro protocols are available \cite{WendzelKeller:CCEng,BacksWendzelKeller:DynRoutingCCs}.

Moreover, the selection of patterns can be done with adaptive techniques \cite{DBLP:conf/prdc/YarochkinDLHK08}. Adaptive covert channels dynamically customize the use of covert channel techniques in order to bypass blocked communications \cite{DBLP:conf/prdc/YarochkinDLHK08} and thus provide a more reliable data transfer. In combination with micro protocols and pattern hopping, adaptive covert channels could not only switch between network protocols but between patterns in case a technique or pattern will be administratively blocked. While it is comparably easy to block a specific covert channel technique, it is more challenging to eliminate a covert channel that switches to a different \emph{pattern} if one pattern was blocked, as this requires implementing countermeasures against multiple patterns.

\section{Countermeasures}\label{Sect:PreventionDetection}

Protection techniques for covert channels either aim on eliminating a covert channel, limiting a channel's capacity or detecting a covert channel \cite{journals/comsur/ZanderAB07}.
The research community considers covert channel protection a \emph{challenging task} \cite{gianvecchio07} and due to the large number of existing covert channel techniques, it is currently difficult to counter \emph{all} covert channels in practice.

Previous approaches only targeted selected covert channels in a given network protocol.
The introduction of patterns which comprise a generic description of a hiding technique enables a more practical approach to develop countermeasures for a whole set of hiding techniques linked to the same pattern.
As our \NumOfPatterns{} patterns represent at least the \NumOfAnalyzedTechniques{} discussed covert channel techniques, a comparably small number of approaches would be enough to counter these covert channel techniques. Thus, the integration of pattern-based countermeasures will lead to a significant reduction of necessary protection mechanisms in practice. 

On the other hand, some specific countermeasures may be more effective than a more general technique that can counter \emph{all} covert channels associated with a particular pattern. For instance, some countermeasures are optimized for detecting hidden data in a header field of a particular protocol. The direct adaption of such a countermeasure to another network protocol, where the particular header field is linked to a different value distribution, may lead to a lower detection accuracy. Therefore, countermeasures for patterns require a \emph{variation} (analog to Sect.~\ref{Sect:Transformation}) as well in order to adapt them to particular network protocols.

\subsection{Countermeasures for Patterns}

In order to evaluate the applicability of existing countermeasures to patterns, we focused on countermeasures covered by the surveys of Llamas \emph{et al.}~\cite{Llamas:Survey} and Zander \emph{et al.}~\cite{journals/comsur/ZanderAB07}. Our evaluation does not include measures that focus on local covert channels (e.g. the \emph{fuzzy time} approach \cite{HuFuzzyTime}) or on the prevention of covert channels at the time of system development (e.g. the \emph{shared resource matrix methodology} or \emph{covert flow trees} \cite{Kemmerer_SharedResMatrix,CovertFlowTrees}). In particular, we considered traffic normalization, the network pump, statistical approaches, and machine learning approaches.

\subsubsection{Traffic Normalization (TN)} Traffic normalizers remove ambiguities and policy-breaking elements in network traffic, , which makes them effective especially against storage channel patterns. The application of normalizers results in side-effects as the normalization of PDU headers often includes setting header fields to default values, i.e. these fields are not usable anymore \cite{NormalisierungsPaper}. Existing traffic normalizers are, for instance, the {\it network-aware active warden} \cite{NetworkAwareActiveWardensIPv6}, the {\it Snort normalizer} \cite{Snort:Normalizer}, and {\it norm} \cite{NormalisierungsPaper} which, taken together, provide more than 100 normalization techniques.
Literature divides normalizers into \emph{stateless} and \emph{stateful} normalizers. Stateless normalizers focus on one packet at a time, and they do not take previous packets into account, while stateful normalizers cache information of previously received packets to evaluate traffic in a more advanced manner and can also detect \emph{more} covert channels, as shown in \cite{springerlink:10.1007/11767831_10}.\\
For always legitimate values (e.g. allowed values for the IPv4 \emph{protocol} field or allowed destination addresses in LAN frames), the creation of normalization rules is challenging and linked to constrictive side-effects \cite{FiskEtAl03}, which makes normalization ineffective against different forms of the \emph{Value Modulation pattern}\footnote{On the other hand, normalizers can eliminate TTL-based covert channels as described in \cite{ZanderEtAl:TTL:Capacity} by setting the TTL value of all packets to the same value.}.\\
A problem of traffic normalizers is their limited buffer size \cite{NormalisierungsPaper}: Buffers cache packets of data flows to re-assemble these flows. Normalization techniques which require large buffers, for instance, to re-order network packets (\emph{PDU Order pattern}) or to normalize the inter-arrival timings and data rate (\emph{Inter-arrival Time pattern and Rate pattern}), are only useful as long as the normalizer's resources are not exhausted. Another problem, especially in IP networks, is that traffic can take different routes and thus, not all packets of a connection pass a normalizer which can result in incomplete information about connections (e.g. missed packets of TCP handshakes \cite{NormalisierungsPaper}).\\
Traffic normalization can only be applied to all network traffic (`blind' normalization) if the normalization is transparent (it does not affect the traffic significantly). Hence, blind normalization cannot eliminate all covert channels. However, if accurate detection methods exist, detected covert channels can be eliminated or limited using targeted normalization or even disruptive measures, e.g. the overt traffic could simply be blocked.\\
%

\subsubsection{Network Pump and Related Concepts (NPRC)} Techniques to limit the capacity of network covert timing channels based on the \emph{Inter-arrival Time} and \emph{Rate patterns} (e.g. the pump \cite{KangMoskowitz:Pump} and the ACK filter \cite{CC_Elimination_Protos}) are traffic normalizers as well. These countermeasures either prevent the entire data flow from HIGH to LOW or do only allow the transmission of acknowledgement messages from HIGH to LOW (related to a data flow from LOW to HIGH).\\
 Another approach presented by Wendzel and Keller \cite{WendzelKeller:PCAW:Journal} limits the covert channel discussed in \cite{wendzel12detecting} (\emph{Value Modulation pattern}) which encodes hidden information in the sequence of utilized network protocols. The approach is called a \emph{protocol switching-aware active warden} (PCAW) and introduces delays on protocol switches.\\
 As in the case of traffic normalization, buffer sizes and routing effects limit the capabilities of the NPRC approaches.

\subsubsection{Statistical Approaches (SA)} The detection of covert timing channels based on inter-arrival times was achieved by Cabuk \emph{et al.} \cite{CabukTimingChanDet_NEU} by representing recorded, rounded inter-arrival times as strings. The strings are compressed and the \emph{compressibility} of a string is used as an indicator for the presence of a covert timing channel. The method takes advantage of the fact that covert timing channels generate traffic with a few characteristic inter-arrival times to signal the different covert bits and thus, result in a few similar strings which can be compressed more efficiently than normal random inter-arrival times. Cabuk \emph{et al.} presented two additional statistical approaches,  one based on the calculation of the $\epsilon$-\emph{similarity} and another based on the standard deviation of recorded inter-arrival times \cite{CabukTimingChanDet_NEU}.\\
Berk \emph{et al.} developed another detection approach for inter-arrival time channels \cite{BerkEtAl}. Their technique uses the fact that timing channels generate inter-arrival time distributions that differ from inter-arrival time distributions of normal application traffic. Gianvecchio and Wang showed that covert timing channels can be detected with high accuracy by analyzing the entropy of inter-arrival times \cite{gianvecchio07}.

While statistical approaches have been effective primarily against channels based on the \emph{Inter-arrival Time pattern} and the \emph{Random Value pattern}, their application to all other patterns is thinkable, since the use of all covert channels leads to changes of statistical value distributions.

\subsubsection{Machine Learning (ML)} Covert channels can be detected using supervised ML approaches where some statistical features are used to characterize covert channels and normal traffic. Classifier models are then trained based on provided examples of features of covert channels and normal traffic. Sohn \emph{et al.} demonstrated that simple covert channels encoded in the IP ID or TCP ISN field  can be discovered with high accuracy by Support Vector Machines (SVMs) \cite{sohn03}. Tumoian \emph{et al.} showed that a neural network can detect
Rutkowska's TCP ISN covert channel \cite{Rutkowska04chaoscommunication} with high accuracy \cite{tumoian05} (both \emph{Random Value pattern}). Zander \emph{et al.} demonstrated that inter-packet timing channels can be detected by C4.5 decision trees trained on several features \cite{zander11stealthier} (\emph{Inter-arrival Time pattern}). Wendzel and Zander showed that C4.5 decision trees also detect simple protocol switching channels (\emph{Value Modulation pattern}) with high accuracy \cite{wendzel12detecting}. 
Besides the mentioned patterns, the application of ML to detect all other patterns appears possible. 


\subsubsection{Applicability of Countermeasures}

Tab.~\ref{tab:AntiCCMeans} summarizes our findings on the applicability of the discussed countermeasures in the context of covert channel patterns. We did not only take existing applications of countermeasures but also potential applications into account, since no pattern-specific implementations are available yet. In general, the prevention of covert channels is always feasible if \emph{all} traffic is blocked but such approaches are not applicable in practice or demand a high-quality covert channel detection. Therefore, Tab.~\ref{tab:AntiCCMeans} does not cover techniques which block the entire traffic.

\begin{table}[!thb]

\tbl{Application of Covert Channel Countermeasures to Patterns\label{tab:AntiCCMeans}}
 { \begin{tabular}{lccc}
   \toprule
   \textbf{ }      & ~\textbf{Elimination}~	  & ~\textbf{Limitation}~ & ~\textbf{Detection}~  \\
   \midrule
   \multicolumn{2}{l}{\textbf{Storage Channel Patterns}} \\
   \hline
    P1. Size
    Modulation		&	  	 	 		&							& SA/ML			 \\
    P2. Sequence	&	  	TN	 		&							& SA/ML			 \\
    P2.a. Position	&	  	TN	 		&							& SA/ML		 \\
    P2.b. Number of
    Elements		&	  	TN 	 		&							& SA/ML			\\
    P3. Add
    Redundancy		&	  	TN 	 		&							& SA/ML			 \\
    P4. PDU
    Corruption/Loss		&	  	TN 			&							&	SA/ML		\\
    P5. Random
    Value		&	  	TN 	 		&							& SA/ML			\\
    P6. Value
    Modulation		&	  			    &   TN (limited),			& SA/ML			\\
    		        &					&   NPRC					& \\
    P6.a. Case		&	  	TN 	 		&							& SA/ML			 \\
    P6.b. LSB		&	  	TN     		&							& SA/ML			\\
    P7. Reserved/Unused
    			&	  	TN 	 		&							& SA/ML			 \\
   \midrule
   \multicolumn{2}{l}{\textbf{Timing Channel Patterns}} \\
   \hline
    P8. Inter-arrival 
    Time			&	  				&	TN (limited), 	    	& SA/ML	\\
    			    &					&   NPRC					& \\
    P9. Rate			&	  				&	TN (limited), 	    	& SA/ML			\\
    			    &					&   NPRC					& \\
    P10. PDU Order	&	  				&	TN (limited)			& SA/ML		\\
    			    &					&   NPRC					& \\
    P11. Re-Transmission
					&	  	   	 	    &							&	SA/ML		\\
   \bottomrule
  \end{tabular}
  }
\end{table}

\subsection{Illustration: Traffic Normalization}

Although traffic normalization was never discussed in the context of covert channel patterns, the existing normalizers already have pattern-oriented capabilities. As dozens of normalization techniques exist, we will discuss only selected ones to show that pattern-oriented countermeasures are feasible.

For instance, existing normalizers can replace the TTL of IPv4 packets with a fixed value to eliminate covert channels. Traffic normalizers can apply the same technique to counter covert channels in the IPv6 \emph{hop limit} field. The technique thus counters the \emph{LSB pattern} even in case of a pattern variation to different network protocols. The LSB pattern could also be applied to the BACnet NPDU \emph{hop count} field --- no new normalization technique must be implemented for the same pattern.


Fisk \emph{et al.} mention general cases for the application of traffic normalization \cite{FiskEtAl03}. For instance, unused fields can be cleared (\emph{Reserved/Unused pattern}); decreasing fields (like the TTL) can be set to fixed values (\emph{LSB pattern} and \emph{Value Modulation pattern}); and derivate fields (i.e. those depending on other fields, like the \emph{Checksum} or the \emph{Internet Header Length} in IPv4) can be re-placed with correct values or the particular packets can be dropped (\emph{PDU Corruption/Loss pattern}). Such general applications of normalization rules can be used in a protocol-independent manner and are thus capable of countering a whole set of hiding techniques which are based on the same pattern.

\subsection{Illustration: Protocol Switching-aware Active Warden}

The aforementioned PCAW \cite{WendzelKeller:PCAW:Journal} introduces delays on protocol switches and thus limits the bitrate of covert channels that signal hidden information through the use of particular network protocols. As shown in \cite{WendzelKeller:PCAW:Journal}, the PCAW cannot only be successfully applied to protocol switching covert channels based on IPv4 but also to building automation networks using BACnet and thus exemplifies the variation of countermeasures as well.

\section{Conclusion}\label{Sect:Concl}

We evaluated \NumOfAnalyzedTechniques{} network covert channel techniques from the last decades and extracted abstract patterns from these techniques. We were able to represent all \NumOfAnalyzedTechniques{} techniques by \NumOfPatterns{} patterns which we arranged in a hierarchical  catalog based on the \emph{Pattern Language Markup Language} (PLML). Moreover, we showed that the majority of these techniques can be reduced to only four different patterns -- evidence that many network covert channel techniques invented represent very similar techniques. 

The pattern catalog will be provided on-line in order to allow the scientific community to modify and extend the covert channel pattern collection in a moderated process. Our catalog eases the novelty evaluation of future covert channel techniques.

We presented the concept of \emph{pattern variation} for covert channels. Pattern variation allows the automatic adaption of a generic hiding technique to different network protocols without requiring a re-implementation of the technique itself. Since covert channels are a dual-use good, we also introduced the pattern-based approaches pattern hopping and pattern combination to improve the throughput and stealthiness of covert channels.

If prevention approaches counter generic patterns instead of hiding techniques, the number of necessary countermeasures is greatly reduced. Under the assumption that future techniques for covert channels will often fall into one of the existing pattern categories, the value of our pattern-based approach increases even further. To this end, the implementation and evaluation of pattern-based countermeasures in practice is considered important future work.

Additional future work will comprise the generation of a PLML-based 
pattern catalog for \emph{local} covert channels and for \emph{payload-based} hiding techniques.

\section*{ACKNOWLEDGEMENTS}

We thank J{\"o}rg Keller as well as Michael Meier and the reading group at Fraunhofer FKIE for suggesting many improvements to previous versions of this paper.
We also thank the referees for their valuable feedback.

\bibliographystyle{ACM-Reference-Format-Journals}
\bibliography{document}

\received{December 2013}{October 2014}{October 2014}


\end{document}